\documentclass[twocolumn,showpacs,prd]{revtex4}
\usepackage{graphicx}%
\usepackage{dcolumn}
\usepackage{amsmath}
\usepackage{latexsym}

\begin{document}

%%%%%%%%%%%%%%%%%%%%
\title{{\bf{\Large  Dual families of non-commutative quantum systems}}}
%%%%%%%%%%%%%%%%%%%%
\author{Frederik G. Scholtz} 
\altaffiliation{fgs@sun.ac.za}
\author{Biswajit Chakraborty} 
\altaffiliation{biswajit@bose.res.in}
\affiliation{Institute of Theoretical Physics, University of
Stellenbosch, Private Bag X1, Matieland 7602, \\
South Africa;}
\affiliation{Satyendra Nath Bose National Centre for Basic Sciences, \\
Block-JD, Sector-III, Salt Lake, Kolkata - 700098, India}
\author{Sunandan Gangopadhyay}
\altaffiliation{sunandan@bose.res.in}
\author{Arindam Ghosh Hazra}
\altaffiliation{arindamg@bose.res.in}
\affiliation{Satyendra Nath Bose National Centre for Basic Sciences, \\
Block-JD, Sector-III, Salt Lake, Kolkata - 700098, India;}

\date{\today}

%%%%%%%%%%%%%%%%%%%%%%%%%%%%%%%%%%%%%%%%%%%%%%%%%%%%%%%%%%%%%%%%%%%%%%%%%%%%
%%%%%%%%%%%%%% abstract
%%%%%%%%%%%%%%%%%%%%%%%%%%%%%%%%%%%%%%%%%%%%%%%%%%%%%%%%%%%%%%%%%%%%
\begin{abstract} 
\noindent
We demonstrate how a one parameter family of interacting non-commuting Hamiltonians, which are physically equivalent, can be constructed in non-commutative quantum mechanics.  This construction is carried out exactly (to all orders in the non-commutative parameter) and analytically in two dimensions for a free particle and a harmonic oscillator moving in a constant magnetic field.  We discuss the significance of the Seiberg-Witten map in this context.  It is shown for the harmonic oscillator potential that an approximate duality, valid in the low energy sector, can be constructed between the interacting commutative and a non-interacting non-commutative Hamiltonian.  This approximation holds to order $1/B$ and is therefore valid in the case of strong magnetic fields and weak Landau-level mixing.
\end{abstract}
\maketitle     
%%%%%%%%%%%%%%%%
%%%%%%%%%%%%%%%%%%%%%%%%%%%%%%%%%%%%%%%%%%%%%%%%%%%%%%%%%%%%%%%%%%
%%%%%%%%%%%%%%%         Introduction
%%%%%%%%%%%%%%%%%%%%%%%%%%%%%%%%%%%%%%%%%%%%%%%%%%%%%%%%%%%%%%%%
\section{Introduction}
Non-commutative quantum field theories \cite{nek} have been
 studied extensively because of its direct connection
 to certain low energy limits of string theory \cite{sw}.
 Non-commutative spaces can arise as brane configurations
 in string theory and in the matrix model of M-theory \cite{suss}.
 Despite the large body of literature on
 non-commutative quantum field theories, 
relatively few works explore the quantum mechanics
 of particles on non-commutative spaces
 \cite{mezincescu, chakraborty, gamboa}.
 
Recently the Seiberg-Witten map has begun to play 
a central role in the analysis of non-commutative 
quantum field theories.  The rational behind this 
map derives from the observation that commutative 
and non-commutative field theories result from 
different regularizations of the same gauge theory,
 at least in two dimensions.  Thus, there should exist
 a map between these theories which reflects the fact
 that the physical content of the two theories is the same.
  The Seiberg-Witten map is therefore a map from the
 non-commutative to the commutative space which preserves
 the gauge invariance and the physics \cite{sw}. However, due to
 its classical nature, it is not clear whether this map will
 hold at the quantum level or not \cite{unknown}.  It is therefore natural to
 enquire about the status of this map in non-commutative
 quantum mechanics where, apart from a few works
 \cite{chakraborty,kokado} which consider the
 Seiberg-Witten map only to lowest order in the 
non-commutative parameter, very little has been done.    
 
 A second motivation for the present work comes from 
the by now well known non-commutative paradigm associated 
with the quantum Hall effect \cite{bigarti,nair}.  
In particular \cite{jelal} explores the possibility 
that the quantum Hall effect in  non-commutative space
 can serve as a paradigm for the fractional quantum Hall
 effect. On the other hand it was realized immediately
 after the discovery of the fractional quantum Hall effect
 that the Coulomb interaction plays an essential role in
 the understanding of this phenomenon \cite{laughlin}.
  This raises the question whether the non-commutative
 Hamiltonian introduced by \cite {jelal} in a somewhat
 ad hoc way can be reinterpreted as an effective
 non-commutative Hamiltonian which describes the same
 physics as the interacting commutative theory, at
 least in some approximation.  Clearly, this equivalence
 cannot be exact as it is well known 
\cite {mezincescu, chakraborty} that a
 non-interacting commutative Hamiltonian with
 constant magnetic field maps onto a non-interacting
 non-commutative Hamiltonian with constant magnetic
 field. However, one might contemplate the possibility
 that there is some preferred value of the non-commutative
 parameter which minimizes the interaction on the
 non-commutative level.  If this is the case the 
corresponding non-interacting non-commutative Hamiltonian
 might be a good starting point for a computation which
 treats the residual interaction as a perturbation.
  This might seem problematic due to the degeneracy 
of the Landau levels.  However, under the assumption of
 a central potential this construction can be carried out
 in each angular momentum sector, which effectively lifts
 this degeneracy and allows for a perturbative treatment
 in each sector (see section \ref{duality}).  
 
Keeping the philosophy of the above remarks in mind,
 i.e. the physical equivalence of different non-commutative
 descriptions, the following question poses itself quite
 naturally: how should a family of non-commutative
 Hamiltonians be parameterized as a function of the
 non-commutative parameter to ensure that they are
 physically equivalent?  This is the central issue
 addressed here. The relation to the Seiberg-Witten
 map and the possible use to construct dualities
 are natural secondary issues that arise which will
 also be addressed, although not in complete generality here.    
 
This paper is organized as follows.  In section
 \ref{general} we consider the general construction
 of a one parameter family of non-commutative, physically
 equivalent Hamiltonians.  In section \ref{free} and
 \ref{harmonic} we apply this general construction to
 a particle in two dimensions moving in a constant
 magnetic field without interactions and in the presence
 of a harmonic potential, respectively.  The construction
 is done to all orders in the non-commutative parameter.
   In section \ref{seiberg-witten} the relation between 
this construction and the Seiberg-Witten map is discussed.
 In section \ref{duality} we construct for an harmonic oscillator potential
 an approximate duality between the interacting commutative
 Hamiltonian and a non-interacting non-commutative Hamiltonian. Section \ref{conclusions} contains our
 discussion and conclusions.  Notational issues are summarized
 in appendix \ref{appendix}.    
%%%%%%%%%%%%%%%%%%%%%%%%%%%%%%%%%%%%%%%%%%%%%%%%%%%%%%%%%%%%%%%%%%%%%%%%%%
%%%%%%%%%%%%%%%%%%% General formalism %%%%%%%%%%%%%%
%%%%%%%%%%%%%%%%%%%%%%%%%%%%%%%%%%%%%%%%%%%%%%%%%%%%%%%%%%%%%%%%%%%%
\section{General considerations}
\label{general}
We consider a non-relativistic particle moving in a plane under a potential $V$ and coupled minimally to a $U(1)$ gauge field $A$.  In commutative space the Hamiltonian reads ($\hbar=c=e=1$)
\begin{eqnarray}
H=\frac{\left({\bf{p}} - \bf{A}\right)^2}{2m}+V(x). 
\label{1}
\end{eqnarray}
To go over to the non-commutative space we replace the commutative quantities by non-commutative ones, denoted by a hat, and introduce the star product, defined in the usual way
\begin{eqnarray}
\left(\hat f\star \hat g\right)(x) = e^{\frac{i}{2}\theta^{\alpha\beta}\partial_{\alpha}\partial^{'}_{\beta}} 
  \hat f(x)\hat g(x^{'})\big{|}_{x^{'}=x} \;
\label{2}
\end{eqnarray}
with $\theta$ the non-commutative parameter. We assume that there is no space time non-commutativity $(\theta^{0i} = 0)$ and, for a planar system, the spatial part of the $\theta$-matrix can be written as $\theta^{ij} = \theta \epsilon^{ij}$.  The Schr\"{o}dinger equation in non-commutative space therefore reads
\begin{eqnarray}
i\frac{\partial \hat\psi({\bf{x}}, t)}{\partial t}
&=&\left[\frac{\left({\bf{p}} - \bf{\hat A}\right)\star \left({\bf{p}} - \bf{\hat A}\right)}{2\hat m}+\hat V(x)\right]\star
\hat\psi({\bf{x}}, t)\nonumber\\
&=&\hat H\star\hat\psi({\bf x}, t)\equiv\hat H_{BS}(\theta)\hat\psi(\theta).
\label{g1}
\end{eqnarray}
Here $\hat H_{BS}(\theta)$ denotes the Hamiltonian after the star product has been replaced by a Bopp-shift, defined by \cite{mezincescu,gamboa,bigarti}
\begin{eqnarray}
\left(\hat f\star \hat g\right)(x) =\hat f\left(x-\frac{\theta}{2}\epsilon^{ij}p_j\right)\hat g(x).
\label{Boppshift}
\end{eqnarray}
Note that the quantities appearing in $\hat H_{BS}(\theta)$ are still the non-commutative ones.
 
The condition that the physics remains invariant under a change in $\theta$ requires that
$\hat H_{BS}(\theta)$ and $\hat H_{BS}(0)$ are related by a unitary transformation,
\begin{eqnarray}
\hat H_{BS}(\theta)=U(\theta)\hat H_{BS}(0)U^\dagger(\theta)\;
\label{g2}
\end{eqnarray}
and that 
\begin{eqnarray}
\hat\psi(\theta)=U(\theta)\hat\psi(0)\;.
\label{g3}
\end{eqnarray}
Differentiating (\ref{g2}) with respect to $\theta$, we obtain
\begin{eqnarray}
\frac{d\hat H_{BS}(\theta)}{d\theta}=[\eta(\theta), \hat H_{BS}(\theta)]\; 
\label{g4}
\end{eqnarray}
where
\begin{eqnarray}
\eta(\theta)=\frac{dU(\theta)}{d\theta}U^\dagger(\theta)
\label{g41}
\end{eqnarray}
is the generator of the unitary transformation relating the non-commutative Bopp-shifted Hamiltonian with the commutative Hamiltonian. 

Let us consider under what conditions eq. (\ref{g4}) admits a solution for $\eta$. These conditions will, of course, provide us with the constraints on the parameterization of the non-commutative Hamiltonian necessary to ensure unitary equivalence, i.e., the existence of $\eta$. It is a simple matter to verify that eq. (\ref{g4}) admits a solution for $\eta$ if and only if
\begin{eqnarray}
\langle n , \theta \vert \frac{d\hat{H}_{BS}(\theta)}{d\theta} \vert n , \theta\rangle = 0\quad ,\quad \forall n
\label{gp5} 
\end{eqnarray}
where $\vert n , \theta\rangle$ are eigenstates of $\hat{{H}}_{BS}(\theta)$, i.e., 
\begin{eqnarray}
\hat{H}_{BS}(\theta) \vert n , \theta\rangle = E_n \vert n , \theta\rangle .
\label{gp4} 
\end{eqnarray}
If eq. (\ref{gp5}) holds, the off-diagonal part of $\eta$ is uniquely determined by 
\begin{eqnarray}
\eta=\sum_{n\neq m}\frac{\langle n,\theta|\frac{d {\hat H}_{BS}}{d\theta}|m,\theta\rangle}{E_m-E_n}|n,\theta\rangle\langle m,\theta|,
\label{eta}
\end{eqnarray}       
while the diagonal part is arbitrary, reflecting the arbitrariness in the phase of the eigenstates.  Here we have assumed no degeneracy in the spectrum of $\hat H_{BS}(\theta)$.  The generalization to the case of degeneracies is straightforward.

The set of conditions (\ref{gp5}) should be viewed as the set of conditions which determines the $\theta$-dependency of the matrix elements of the non-commutative potential $\hat V$ and gauge field $\hat A$.  Clearly one would expect that these matrix elements are under-determined, i.e., that not both $\hat V$ and $\hat A$ are uniquely determined by them.  Instead one can fix one of these and compute the other.  For comparison with the Seiberg-Witten map it is therefore natural to take for $\hat A$ the non-commutative gauge field as determined from the Seiberg-Witten map. Note that this procedure implies that $\hat V$ will be gauge dependent. 

Consider the Seiberg-Witten map for the non-commutative wave-function which reads to lowest order in $\theta$ \cite{bichl} $\hat\psi=\psi-\frac{1}{2}\theta\epsilon^{ij}A_i\partial_j\psi$. Below we consider two dimensional systems in a constant magnetic field.  Taking the symmetric gauge, the Seiberg-Witten map reduces to a $\theta$ dependent scaling transformation. Clearly this is not a unitary transformation and a unitary Seiberg-Witten map can be constructed as in \cite{kokado}.  However, a more convenient point of view, closer in spirit to the Seiberg-Witten map, would be to relax the condition of unitarity above.  It therefore seems worthwhile, in particular to relate to the Seiberg-Witten map, to generalize the above considerations by relaxing the condition of unitarity.   

This generalization is straightforward.  We simply have to replace the unitary transformation in (\ref{g2}) and (\ref{g3}) by a general similarity transformation
\begin{eqnarray}
\hat H_{BS}(\theta)=S(\theta)\hat H_{BS}(0)S^{-1}(\theta)\;,
\label{g2a}
\end{eqnarray}
while
\begin{eqnarray}
\hat\psi(\theta)=S(\theta)\hat\psi(0),
\label{g3a}
\end{eqnarray}
and note that a new inner product $\langle \psi|\phi\rangle_T=\langle \psi|T|\phi\rangle$ can be defined such that $\hat H_{BS}(\theta)$ is hermitian with respect to it. In particular $T$ is given by $T=(S^{-1})^{\dagger} S^{-1}$ and has the property $T\hat H_{BS}(\theta)=\hat H_{BS}^\dagger(\theta)T$.  Under this prescription the same physics results.  A detailed exposition of these issues can be found in \cite{scholtz}. 

Differentiating (\ref{g2a}) with respect to $\theta$, we obtain
\begin{eqnarray}
\frac{d\hat H_{BS}(\theta)}{d\theta}=[\eta(\theta), \hat H_{BS}(\theta)]\;
\label{g4a}
\end{eqnarray}
where 
\begin{eqnarray}
\eta(\theta)=\frac{dS(\theta)}{d\theta}S^{-1}(\theta)
\label{g44a}
\end{eqnarray}
 is now the generator of the similarity transformation relating the non-commutative Bopp-shifted Hamiltonian with the commutative Hamiltonian. 

It is now a simple matter to verify that (\ref{gp5}) gets replaced by
\begin{eqnarray}
\langle n , \theta \vert T\frac{d\hat{H}_{BS}(\theta)}{d\theta} \vert n , \theta\rangle = 0\quad ,\quad \forall n
\label{gp5a} 
\end{eqnarray}
where $\vert n , \theta\rangle$ are eigenstates of $\hat {H}_{BS}(\theta)$ (note that the eigenvalues will be real as $\hat {H}_{BS}(0)$ is assumed hermitian and thus has real eigenvalues). As before, if eq. (\ref{gp5a}) holds the off-diagonal part of $\eta$ is uniquely determined by 
\begin{eqnarray}
\eta=\sum_{n\neq m}\frac{\langle n,\theta|T\frac{d \hat H_{BS}}{d\theta}|m,\theta\rangle}{E_m-E_n}|n,\theta\rangle\langle m,\theta|T
\label{eta10}
\end{eqnarray}       
while the diagonal part is arbitrary, reflecting the arbitrariness in the phase and now also the normalization of the eigenstates. 

Under the above description, the Hamiltonians $\hat {H}_{BS}(\theta)$ and $\hat {H}_{BS}(0)$ are physically equivalent.  There is, however, one situation in which this equivalence may break down and of which careful note should be taken.  This happens when the similarity transformation $S(\theta)$ becomes singular for some value of $\theta$, which will be reflected in the appearance of zero norm or unnormalizable states in the new inner product.  Only values of $\theta$ which can be reached by integrating (\ref{g44a}) from $\theta=0$ without passing through a singularity, can be considered physically equivalent to the commutative system. 
  
To solve (\ref{gp5}) or (\ref{gp5a}) in general is of course impossible. Therefore we take a slightly different approach in what follows.  We take an ansatz for $\eta$ and solve (\ref{g4}) or (\ref{g4a}) directly.  The ansatz for $\eta$ is again motivated by the Seiberg-Witten map.  We have already noted above that in the cases of interest to us, i.e., two dimensional systems in constant magnetic fields, the Seiberg-Witten map for the non-commutative wave-function corresponds to a scaling transformation.  Motivated by this we make the following ansatz             
\begin{eqnarray}
\eta(\theta)= f(\theta)r\partial_{r}=if(\theta)x.p
\label{g5}
\end{eqnarray}
with $f$ an arbitrary function to be determined.  The finite form of this scaling transformation is simply obtained by integrating (\ref{g44a}) to yield
\begin{eqnarray}
S(\theta)= e^{i\left(\int_0^\theta f(\theta^{\prime})d\theta^{\prime}\right)x.p}.
\label{scaletrans}
\end{eqnarray}
Clearly this is not a unitary transformation and therefore falls in the class of more general transformations described above.  Furthermore we note that the non-singularity of $S(\theta)$ requires that the integral $\int_0^\theta f(\theta^{\prime})d\theta^{\prime}$ exists.
%%%%%%%%%%%%%%%%%%%%%%%%%%%%%%%%%%%%%%%%%%%%%%%%%%%%%%%%%%%%%%%%%%%%%
%%%%%%%%         Free particle in a constant magnetic field
%%%%%%%%%%%%%%%%%%%%%%%%%%%%%%%%%%%%%%%%%%%%%%%%%%%%%%%%%%%%%%%%%%%%
\section{Free particle in a constant magnetic field}
\label{free}
In this section we apply the considerations discussed above to the case of a free particle ($\hat V=0$) moving in a non-commutative plane in the presence of a constant non-commutative magnetic field.  The Schr\"{o}dinger equation is given by (\ref{g1}) with $\hat V$ set to zero. 

In the symmetric gauge $\hat A_{i}=-\frac{\bar B(\theta)}{2}\epsilon_{ij}x_{j}$ \footnote{We use $\bar B(\theta)$ to denote the non-commutative counterpart of $B$
in (\ref{symgauge}). It should not be confused with the non-commutative magnetic field $\hat B$ as determined from the field strength (see (\ref{s2})).  In the limit $\theta=0$, $\bar B(\theta)=B$.} the Bopp-shifted Hamiltonian is easily found to be 
\begin{eqnarray}
 \hat H_{BS}(\theta) &=& \frac{\left(1 + \frac{\bar B\theta}{4}\right)^2}{2\hat m(\theta)}\left({\bf{p}} - \frac{1}{1 + \frac{\bar B\theta}{4}}{\bf A}\right)^2 \nonumber\\
&=& \frac{1}{2M(\theta)}\left(p^{2}_{x} + p^{2}_{y}\right) + \frac{1}{2}M(\theta){\Omega}(\theta)^{2}\left( x^2 + y^2 \right)\nonumber\\ &&- \Omega(\theta) L_{z}\; 
\label{5} 
\end{eqnarray}
where
\begin{eqnarray} 
\frac{1}{2M(\theta)} = \frac{\left(1 + \frac{\bar B\theta}{4}\right)^2}{2\hat m(\theta)}\quad,\quad \frac{1}{2}M(\theta){\Omega(\theta)}^{2} = \frac{\bar B^2}{8\hat m(\theta)}\;.
\label{freepar}
\end{eqnarray}
Substituting the above form of the Hamiltonian in (\ref{g4}) with $\eta$ as in (\ref{g5}), we obtain
the following set of differential equations:
\begin{eqnarray}
\frac{d M^{-1}(\theta)}{d\theta} &=& -2f(\theta)M^{-1}(\theta)\;, \\  
\frac{d \left(M(\theta){\Omega(\theta)}^{2}\right)}{d\theta} &=& 2M(\theta){\Omega(\theta)}^{2}f(\theta)\;,
\label{6a} 
\end{eqnarray}
\begin{eqnarray}
\frac{d\Omega(\theta)}{d\theta} &=& 0.
\label{5a} 
\end{eqnarray}
Note that (\ref{5a}) ensures the stability of the energy spectrum, i.e
the cyclotron frequency $\Omega(\theta)=\Omega(\theta=0)=B/2m$, where
$m=\hat m(\theta=0)$.
This is the physical input in our analysis and will play a very
important role as we shall see later. Note that (\ref{5a}) follows trivially
by combining the equations in (\ref{6a}).
The above equations (\ref{freepar}, \ref{6a}, \ref{5a}) immediately lead to
\begin{eqnarray}
f(\theta) = \frac{1}{2M(\theta)}\frac{dM(\theta)}{d\theta} = \frac{\partial_{\theta}\bar{B}(\theta) - \frac{\bar{B}(\theta)^2}{4}}{2\bar{B}(\theta)\left(1 + \frac{\theta\bar{B}(\theta)}{4}\right)}\;
\label{7a} 
\end{eqnarray}
which fixes $f$ once $\bar B$ has been determined.
  As indicated before, we take $\hat A$
 as the non-commutative gauge field determined from the Seiberg-Witten map.
  With this in mind we now proceed to determine $\bar B$.
  The Seiberg-Witten  maps for $\hat{\psi}$ and  $\hat{A_k}$
 to lowest order in $\theta$ are \cite{bichl,sw}
\begin{eqnarray}
\label{swmap}
\hat{\psi}&=&{\psi}-\frac{1}{2}\theta\epsilon^{ij}A_i\partial_j\psi\;,\nonumber\\
\hat{A_k}&=&{A_k}-\frac{1}{2}\theta\epsilon^{ij}A_i(\partial_jA_k+F_{jk})\;.
\end{eqnarray}

From the Seiberg-Witten transformation of the commutative gauge field $A_k$ to the non-commutative one $\hat{A}_k$, one can easily see that a symmetric gauge configuration 
\begin{equation}
\label{symgauge}
A_i=-\frac{B}{2}\epsilon_{ij}x^j 
\end{equation}
with magnetic field  $B=F_{12}=(\partial_1A_2-\partial_2A_1)$, transforms to a symmetric gauge field configuration at the non-commutative level. Using the same notation as in (\ref{symgauge}) we write
\begin{equation}
\hat{A}_i=-\frac{\bar{B}}{2}\epsilon_{ij}x^j
\label{symgauge1}
\end{equation}
where $\bar B$ is determined to leading order in $\theta$ from (\ref{swmap}) to be
\begin{equation}
\bar{B}=B(1+\frac{3\theta B}{4}).
\label{solution}
\end{equation}
As pointed out in [15], $\bar{B}$ should not be identified with the non-commutative magnetic field $\hat{B}$, which has an additional Moyal bracket term $[\hat{A}_1, \hat{A}_2 ]_{\star}$:
\begin{equation}
\hat{B}=\hat{F}_{12} = \partial_1\hat{A}_2-\partial_2\hat{A}_1 - i(\hat{A}_1{\star}\hat{A}_2-\hat{A}_2{\star}\hat{A}_1)=\bar B(1+\frac{\theta\bar B}{4}).
\label{s2}
\end{equation}
This is precisely the same expression one gets if one applies the Seiberg-Witten map directly at the level of the field strength tensor, which is given by \cite{sw}:
\begin{equation} 
\hat{F}_{\mu\nu}=F_{\mu\nu}+\theta\epsilon^{ij}F_{\mu i}F_{\nu j}\;.
\label{s3}
\end{equation}
Note that the expression (\ref{s2}) relating $\hat B$ with $\bar B$ is an exact one in contrast with (\ref{solution}) which relates $\bar B$ to $B$ only up to leading order in $\theta$. For a constant field configuration, the Seiberg-Witten equation for the field strength tensor can be integrated exactly to give the result \cite{sw}
\begin{equation}
\hat{B}=\frac{1}{1 - \theta B}B.
\label{s4}
\end{equation}
From (\ref{s2}) and (\ref{s4}), we obtain a quadratic equation in $\bar B(\theta)$ that can be solved exactly to give
\begin{equation}
\bar B(\theta)=\frac{2}{\theta }\left[(1 - \theta B)^{-1/2} - 1\right].
\label{s5}
\end{equation}
The above expression for $\bar B(\theta)$ is exact up to all orders in $\theta$. When substituted in (\ref{symgauge1}) an expression, correct to all orders in $\theta$, for the non-commutative gauge field $\hat A_{i}$ result 
\begin{equation}
\hat A_{i} = -\frac{1}{\theta }\left[(1 - \theta B)^{-1/2} - 1\right]
\epsilon_{ij}x^{j}.
\label{s6}
\end{equation}
Substituting $\bar B$ from (\ref{s5}) into (\ref{7a}) yields
\begin{equation}
f(\theta)=\frac{\bar B(\theta)}{4}\;.
\label{f}
\end{equation}
Upon differentiating (\ref{g3a}) with respect to $\theta$ and using $f$ from (\ref{f}), we find that $\hat \psi(\theta)$ must satisfy the following equation:
\begin{eqnarray}
\frac{d{\hat\psi(\theta)}}{d\theta} = \frac{\bar B(\theta)}{4}r\frac{d\hat \psi(\theta)}{dr}\;. 
\label{swmap20} 
\end{eqnarray}
This result can now be compared to the corresponding Seiberg-Witten transformation rule for $\hat\psi$.  The first of the Seiberg-Witten equations listed in (\ref{swmap}) for an arbitrary $\theta + \delta\theta$ reads, 
\begin{eqnarray}
\hat\psi(\theta + \delta\theta)- \hat\psi(\theta) = -\frac{1}{2}\theta\epsilon^{ij}\hat A_i\star\partial_j\hat\psi(\theta).
\label{swmap10} 
\end{eqnarray}
Upon substituting $\hat A_i$ from (\ref{symgauge1}), eq. (\ref{swmap20}) indeed results.  Thus the transformation rule as obtained from the requirement of physical equivalence agrees with that of the Seiberg-Witten map.
 
Finally, substituting $\bar B(\theta)$ in the condition $\Omega=B/2m$ yields the following expression for $\hat m(\theta)$,
\begin{eqnarray}
\hat m(\theta)=\frac{m}{1-\theta B}\;.
\label{1000d} 
\end{eqnarray}
The above equation relates the non-commutative mass $\hat m(\theta)$ with the commutative mass $m$. This generalizes the result obtained in \cite{chakraborty} to all orders in $\theta$.

In a simple case such as this, one can of course solve the Schr\"odinger equation exactly.  It is useful to see what the above procedure entails from this point of view.  To solve for the eigenvalues and eigenfunctions of (\ref{5}) is a standard procedure and for notational completeness we summarize the essential steps in appendix \ref{appendix}.  This results in the degenerate eigenvalue spectrum
\begin{eqnarray}
E_{n_-,\ell}=2\Omega\left( n_{-} + \frac{1}{2}\right),\nonumber\\ n_-=0,1\ldots\quad;\;\ell=-n_-,-n_-+1,\ldots\;,
\label{10.1} 
\end{eqnarray}
where $\ell$ denotes the eigenvalues of the angular momentum operator $L_3$.  The corresponding eigenstates are obtained by acting with the creation operators $b_{\pm}^\dagger$ defined in (\ref{10b}) on the ground-state
\begin{eqnarray}
\hat\psi(z, \bar z;\theta) &=& N\exp\left[-\frac{M\Omega}{2}\bar{z}z\right]\nonumber\\
 &=&  N\exp\left[- \frac{\bar B(\theta)}{4\left(1 + \frac{\bar B(\theta)\theta}{4}\right)}\bar{z}z\right]\;.
\label{10c.1} 
\end{eqnarray}
Comparing with our previous results, we note that eq. (\ref{5a}) ensures invariance of the spectrum under a change of $\theta$.  Furthermore direct inspection shows that the {\it unnormalised} ground-state and, subsequently, also all excited states satisfy the transformation rule (\ref{swmap20}).  The fact that the unnormalised wave functions satisfy the transformation rule (\ref{swmap20}) is consistent with our earlier remarks on the non-unitary nature of the scaling transformation.

Finally, note that although the non-commutative parameters $\bar B(\theta)$ and $\hat m(\theta)$ have singularities at $\theta=1/B$, these singularities cancel in the parameter $\Omega$, which is by construction free of any singularities, i.e., the spectrum is not affected by this singularity.  This is also reflected by the fact that the integral of $f$, as determined in (\ref{f}), is free of this singularity. Thus, despite the appearance of this singularity in the parameters of the non-commutative Hamiltonian, there is no breakdown of the physical equivalence (see the discusion in section \ref{general}).

%%%%%%%%%%%%%%%%%%%%%%%%%%%%%%%%%%%%%%%5%%%%%%%%%%%%%%%%%%%%%%%%%%
%   Potential
%
%%%%%%%%%%%%%%%%%%%%%%%%%%%%%%%%%%%%%%%%%%%%%%%%%%%%%%%%%
%---------------------------------------
\section{Harmonic oscillator in a constant magnetic field}
\label{harmonic}
%--------------------------------------- 
In this section we include a harmonic oscillator potential $V=\lambda r^2$ in the commutative Hamiltonian (\ref{1}).  If the physical equivalence between the non-commutative and commutative Hamiltonians is indeed implementable through a scale transformation, we expect the potential to be form preserving (this is certainly not true for arbitrary potentials). We therefore extend to the non-commutative Hamiltonian by taking for the non-commutative potential in (\ref{g1}) $\hat V=\hat\lambda(\theta)r^2$, where the oscillator strength $\hat\lambda(\theta)$ has to be determined.  Obviously we must also require that $\hat{\lambda}(\theta) = \lambda $ in the limit $\theta = 0$. Taking this form for the non-commutative Hamiltonian (\ref{g1}), the Bopp-shifted Hamiltonian is easily found to be
\begin{eqnarray}
 \hat H_{BS}(\theta) &=& \frac{\left(1 + \frac{\bar B\theta}{4}\right)^2}{2\hat m}\left({\bf{p}} - \frac{1}{1 + \frac{\bar B\theta}{4}}{\bf A}\right)^2 \nonumber\\
 &&+\hat{\lambda}(\theta)\left[\frac{{\theta}^2}{4}\left(p^{2}_{x} + p^{2}_{y}\right) + \left( x^2 + y^2\right) - \theta L_z \right]\nonumber\\
&=& \frac{1}{2M}\left(p^{2}_{x} + p^{2}_{y}\right) + \frac{1}{2}M{\Omega}^{2}\left( x^2 + y^2 \right)\nonumber\\
&&- \Lambda(\theta)L_{z}\;
\label{101d} 
\end{eqnarray}
where
\begin{eqnarray}
\label{harpar}
\frac{1}{2M} &=& \frac{\left(1 + \frac{\bar B\theta}{4}\right)^2}{2\hat m} + \frac{\hat{\lambda}{\theta}^2}{4}\; ,\nonumber\\
\frac{1}{2}M{\Omega}^{2} &=& \frac{\bar B(\theta)^2}{8\hat m(\theta)} + \hat{\lambda}(\theta)\;,\\
\Lambda(\theta) &=& \left[\frac{M{\Omega}^2\theta}{2} + \frac{\bar B\left[1 - \left({\frac{M\Omega \theta}{2}}\right)^2 \right]}{2\left(1 + \frac{\bar B\theta}{2}\right)M} \right]\nonumber\;.
\end{eqnarray}
Here $\bar B(\theta)$ is again taken from the Seiberg-Witten map (\ref{s5}). Substituting the above form of the Hamiltonian in (\ref{g4}) with $\eta$ as in (\ref{g5}), we obtain
the following set of differential equations:
\begin{eqnarray}
\frac{dM^{-1}(\theta)}{d\theta} &=& -2f(\theta)M^{-1}(\theta)\;,\\     
\frac{d \left(M(\theta){\Omega(\theta)}^{2}\right)}{d\theta} &=& 2M(\theta){\Omega(\theta)}^{2}f(\theta)\;,
\label{61a} 
\end{eqnarray}
\begin{eqnarray}
\frac{d\Lambda(\theta)}{d\theta} &=& 0\;.
\label{51a} 
\end{eqnarray}
Eq. (\ref{51a}) requires that $\Lambda(\theta)$ is independent of $\theta$ and hence we have the condition $\Lambda(\theta)=\Lambda(0)=B/2m$. Substituting the forms of $M(\theta)$ in terms of $\hat m(\theta)$ and $\hat \lambda(\theta)$, we obtain the following solution for $\hat m(\theta)$ in terms of $\hat \lambda(\theta)$:
\begin{eqnarray}
\hat{m}(\theta) = \frac{m}{\left(1 - \theta B\right)}\frac{B}{\left(B - 2m\theta\hat{\lambda}(\theta)\right)}\;.
\label{600a} 
\end{eqnarray}
The set of differential equations in (\ref{61a}) can also be combined to obtain
\begin{eqnarray}
\frac{d\Omega^2}{d\theta} = 0\;.
\label{601a} 
\end{eqnarray}
This shows that $\Omega$ is a constant and therefore we have 
\begin{eqnarray}
\Omega^2(\theta) = \Omega^2(\theta = 0) = \frac{B^2}{4m^2} + \frac{2\lambda}{m}\;.
\label{602a} 
\end{eqnarray}
Substituting $\Omega^2(\theta)$ in (\ref{harpar}) and using
(\ref{600a}), we get a quadratic equation for $\hat \lambda(\theta)$,
\begin{eqnarray}
\left[ B^3 + 8(1 - \theta B)mB\hat{\lambda}(\theta) - 16(1 - \theta B)m^2\theta \hat{\lambda}(\theta)^2\right]\nonumber\\= B^3 + 8\lambda mB\;.
\label{603a} 
\end{eqnarray}
Solving for $\hat\lambda$ yields
\begin{eqnarray}
\hat{\lambda}(\theta) = \frac{B}{4m\theta}\left[1 - \left(1 - \frac{8\lambda m\theta}{B(1 - \theta B)}\right)^{\frac{1}{2}}\right]\;,
\label{604a} 
\end{eqnarray}
where we have taken the negative sign before the square root since with this choice we have $\hat\lambda(\theta=0)=\lambda$. 

With the value of $\bar B(\theta)$ fixed from the Seiberg-Witten map and $\hat m(\theta)$ and $\hat\lambda(\theta)$ determined as above, we can compute the value of $M(\theta)$ from (\ref{harpar}) and subsequently the value of $f(\theta)$ from (\ref{61a}) as $f(\theta)=\frac{1}{2M(\theta)}\frac{dM(\theta)}{d\theta}$.  This is a rather lenghthy expression which we do not need to list for our present purposes. What is important to note, however, is that once $f(\theta)$ is fixed, the transformation rule satisfied by $\hat\psi(\theta)$ is determined from (\ref{g3a}) and that this transformation rule is not the same as the one derived from the Seiberg-Witten map (\ref{swmap20}).  In fact, the transformation rule for $\hat\lambda(\theta)$ also turns out to be different from the Seiberg-Witten map.  We discuss these points in more detail in the next section.

As a consistency check one can once again solve for the eigenvalues and eigenstates.  The procedure is the same as in appendix \ref{appendix} and one finds for the eigenvalues
\begin{eqnarray}
 E_{n_-,\ell}= 2\Omega \left(n_{-} + \frac{1}{2}\right) +(\Omega- \Lambda)\ell\;,\nonumber \\
 n_-=0,1\ldots;\;\ell=-n_-,-n_-+1,\ldots\;.
\label{z2} 
\end{eqnarray}
Note that the degeneracy in $\ell$ has been lifted. However, in the limit $\lambda = 0$, the energy spectrum given by (\ref{10.1}) is recovered.  The corresponding eigenstates are again obtained by acting with the creation operators $b_{\pm}^\dagger$ defined in (\ref{10b}) on the ground-state
\begin{eqnarray}
\hat\psi(z, \bar z; \theta) = N\exp\left[-\frac{1}{4}\sqrt{\frac{2\bar B(\theta)^2 + 16 \hat
\lambda \hat m}{2\left(1+\frac{\theta \bar B(\theta)}{4}\right)^2+ 
\theta^2 \hat\lambda \hat m}}\bar{z}z\right]\;.\nonumber\\
\label{z3} 
\end{eqnarray}
Once again we note that (\ref{61a}) and (\ref{51a}) ensures invariance of the spectrum under a change in $\theta$.  Using the values of $\bar B(\theta)$, $\hat m(\theta)$ and $\hat\lambda(\theta)$ as determined above, one finds that the {\it unnormalised} wave-functions indeed satisfy the transformation rule as determined by (\ref{g3a}) and not the Seiberg-Witten transformation rule (\ref{swmap20}).
Also, in the $\theta=0$ limit, (\ref{z3}) smoothly goes over to the standard 
commutative result
\begin{eqnarray}
\hat\psi(z, \bar z, \theta=0) = \psi(z, \bar z)= N\exp\left[-\frac{1}{4}\sqrt{B^2+ 8\lambda m }\bar{z}z\right]\;.\nonumber\\
\label{z300} 
\end{eqnarray}

Finally we remark on the non-singularity of the scaling transformation $S(\theta)$.  As already pointed out in section \ref{general} this requires the existence of the integral of $f$, which in the present case is simply given by $\log(M(\theta)/m)/2$.  This turns out to be free of singularities, although the non-commutative parameters again exhibit singularities at $\theta=1/B$.  As in the free case these singularities cancel in the parameters $\Omega$ and $\Lambda$ which determine the physical spectrum. 
%%%%%%%%%%%%%%%%%%%%%%%%%%%%%%%%%%%%%%%%%%%%%%%%%%%%%%%%%%%%%%%%%%
%    Relation to Seiberg-Witten map
%%%%%%%%%%%%%%%%%%%%%%%%%%%%%%%%%%%%%%%%%

\section{Relation to Seiberg-Witten map}
\label{seiberg-witten}
In this section, we are going to discuss the relationship of the flow
equations for $\hat m$ and $\hat \lambda$ obtained from the stability
analysis of the previous section to the flow equation obtained from
the Seiberg-Witten map. To that end, let us write down the $U(1)_{\star}$ gauge
invariant action from which the ${\star}$ gauge covariant 
one-particle Schr\"{o}dinger
equation (\ref{g1}) follows as Euler-Lagrangian equation:
\begin{eqnarray}
\hat {S}=\int d^3{x} \hat{\psi}^{\dagger}\star(i\hat{D}_0+\frac{1}
{2\hat {m}}\hat{D}_i\star\hat{D}_i+\hat{V})\star\hat{\psi}\;.
\label{p420} 
\end{eqnarray}

The preservation of $U(1)_\star$ gauge invariance of the action requires 
that the potential $\hat{V}$ must transform adjointly under $\star$ gauge transformation
\begin{eqnarray}
\hat{V}(x)\longrightarrow\hat{V}^\prime(x)=\hat{U}(x){\star}\hat{V}(x){\star}\hat{U}^{\dagger}(x)
\label{p4} 
\end{eqnarray}
for $\hat U(x)\in U(1)_{\star}$.
The reason for this is quite simple to see.  If it were to remain invariant, this would have implied that the
Moyal bracket between $\hat{V}$ and $\hat{U}$, $\forall$ $\hat U \in  U(1)_{\star}$ vanishes $([\hat{V}, \hat{U}]_{\star}=0)$.  Through Wigner-Weyl correspondence this in turn implies that $V_{\rm op}$ commutes with $U_{\rm op}$ at the operator level:  $[V_{\rm op}, U_{\rm op}]=0\; \forall U_{\rm op}$.  Applying Schur's lemma, assuming that $U_{\rm op}$ acts irreducibly, this indicates $V_{\rm op}=$constant.
Clearly this does not have the desired property. Now the Seiberg-Witten
 transformation property of $\hat{V}(x)$ can be easily obtained as 
\begin{eqnarray}
\hat{V}^{\prime}(x) = \hat{V}(x) - \delta\theta\epsilon^{ij}\hat{A}_i\star\partial_j\hat{V}(x)\;,
\label{p5} 
\end{eqnarray}
which relates the non-commutative potential $\hat{V}(x;\theta) \equiv \hat{V}(x)$ for non-commutative parameter $\theta$ to the corresponding non-commutative potential $\hat{V}(x;\theta + \delta\theta) \equiv \hat{V}^{\prime}(x)$ for non-commutative parameter $(\theta + \delta\theta)$. For the non-commutative gauge potential (\ref{symgauge1}), this leads to the following differential equation 
\begin{eqnarray}
\frac{d{\hat V(\theta)}}{d\theta} = \frac{\bar B(\theta)}{2}r\frac{d\hat V(\theta)}{dr} 
\label{p6} 
\end{eqnarray}
which can be solved by the method of seperation of variables \footnote{Such a seperation of variables can be made as one can expect that
a commutative central potential goes over to another central potential
at the non-commutative level.}, i.e. by taking $\hat V(r,\theta) = V(r)\hat\lambda_{sw}(\theta)$. We also have the boundary condition $\hat\lambda_{sw}(\theta=0)=\lambda$.
Using this, (\ref{p6}) simplifies to 
\begin{eqnarray}
\frac{2}{\bar B(\theta)\hat\lambda_{sw}(\theta)}\frac{d\hat\lambda_{sw}(\theta)}{d\theta} = \frac{r}{V(r)}\frac{dV(r)}{dr} = k(=\mathrm{constant}).\nonumber\\
\label{z9} 
\end{eqnarray}
Solving we get 
\begin{eqnarray}
V(r) &=& \lambda r^k\;, \nonumber \\
\hat\lambda_{sw}(\theta) &=& \lambda\exp\left[\frac{k}{2}\int^{\theta}_{0}d\theta^{'}\bar B(\theta^{'})\right]\nonumber \\
&=& \lambda\left(\frac{1 + \left(1 - \theta B\right)^{\frac{1}{2}}}{2}\right)^{-2k}\;.
\label{z10} 
\end{eqnarray}
For $k = 2$, we get the usual harmonic oscillator i.e.
\begin{eqnarray}
\hat V(r,\theta) &=& \hat\lambda_{sw}(\theta){r}^{2}\nonumber \\
                 &=& \lambda\left(\frac{1 + \left(1 - \theta B\right)^{\frac{1}{2}}}{2}\right)^{-4}{r}^{2}\;. 
\label{z11}
\end{eqnarray}
If we now demand as in the free case that (\ref{z3}) satisfies (\ref{swmap20})
then the solution of (\ref{swmap20}) can also be found by taking the trial solution $\hat\psi(z,\bar{z};\theta) = N\exp\left(- \frac{\bar{z}z}{4} g(\theta)\right)$ subject to the boundary condition (\ref{z300}) at $\theta = 0 $. This leads to the solution
\begin{eqnarray}
\hat\psi_{sw}(z, \bar z; \theta) = N\exp\left[-\bar{z}z\frac{\sqrt{\left(B^2 + 8m\lambda \right)}}{\left((1 - \theta B)^{\frac{1}{2}} + 1\right)^2}\right]\;.
\label{z5} 
\end{eqnarray}
Comparing equations (\ref{z3}) and (\ref{z5}), we get an algebraic equation,
\begin{eqnarray}
\frac{4\left(B^2 + 8m\lambda \right)^{\frac{1}{2}}}{\left[\left(1 - \theta B\right)^{\frac{1}{2}} +1\right]^2} = \left[\frac{2\bar{B}^{2}(\theta) + 16\hat\lambda_{sw}\hat m_{sw} }{2\left(1 + \frac{\theta\bar{B}(\theta)}{4}\right)^2 + \theta^2 \hat\lambda_{sw}\hat m_{sw}}\right]^{\frac{1}{2}}\;
\label{z6} 
\end{eqnarray}
which leads to
\begin{eqnarray}
&&\hat\lambda_{sw}\hat m_{sw} =\nonumber\\
&&\frac{4m\lambda\left[1 + \left(1 - \theta B\right)^{\frac{1}{2}}\right]^2}{\left(1 - \theta B\right)\left(\{\left(1 - \theta B\right)^{\frac{1}{2}} + 1\}^4 - \theta^{2}\left(B^2 + 8m\lambda \right)\right)}\;.\nonumber\\
\label{z7} 
\end{eqnarray}
Substituting the value of $\hat\lambda_{sw}$ from (\ref{z10}) we obtain the value of $\hat m_{sw}$ as
\begin{eqnarray}
\hat m_{sw} = \frac{m}{4\left(1 - \theta B\right)}\frac{\left[1 + \left(1 - \theta B\right)^{\frac{1}{2}}\right]^6}{\left[\{\left(1 - \theta B\right)^{\frac{1}{2}} + 1\}^4 - \theta^{2}\left(B^2 + 8m\lambda\right)\right]}\;.\nonumber\\
\label{z12} 
\end{eqnarray}
The flow structure of $\hat\lambda_{sw}$ (\ref{z10}) and 
$\hat m_{sw}$ (\ref{z12}) in $\theta$ shows that the Seiberg-Witten flow is different
(in the presence of interactions) from the flows obtained
in the previous section (\ref{600a}) and (\ref{604a}) from
the consideration of the stability of the spectrum, although the formal
structure of the wave-functions $\hat\psi_{sw}$ (\ref{z5}) and $\hat\psi$
(\ref{z3}) are the same.  Indeed, it can be checked easily and explicitly that the flow obtained here ((\ref{z10}) and (\ref{z12})) from the Seiberg-Witten map is not spectrum preserving, as is the case with the flow of the previous section.  This indicates that these flows are not equivalent or related in some simple way. 

We have already seen that in absence of interaction $(\hat\lambda=0)$ the non-commutative wave-function $\hat\psi_{sw}$ satisfies the Seiberg-Witten map, subject to the boundary condition (\ref{10d}) at $\theta = 0$, when $\hat{\psi}_{sw}$ becomes identifiable with the commutative wave-function $\psi$. Also, unlike its non-commutative counterpart $\hat{\psi}$,
the commutative wave-function $\psi$ does not have a flow of its own in $\theta$. However, the situation changes drastically in the presence of interactions. To see this more clearly, let us consider the Schr\"odinger equation,
\begin{eqnarray}
iD_0\psi = -\frac{1}{2m}D_iD_i\psi - \frac{i\theta}{2}\epsilon^{ij}F_{i0}D_j\psi\;
\label{p3} 
\end{eqnarray}
obtained from the $U(1)$ gauge invariant effective action in the presence of a background gauge field, derived in \cite {chakraborty} to leading order in the non-commutative parameter $\theta$. Note that the temporal component $A_0$ of the background gauge field can be regarded as $(-V)$, where $V$ is the potential since this background gauge field is time independent. Indeed the Seiberg-Witten transformation property of both $A_0$ and $V$ become identical, as can be seen from  (\ref{p5}) and (\ref{swmap}). This helps us to identify, again to leading order in $\theta$, the corresponding Hamiltonian as
\begin{eqnarray}
H = \frac{\left({\bf{p}} - \bf{A}\right)^2}{2m} + V - \frac{\theta}{2}\epsilon^{ij}\partial_iV\left({p_j} -  A_j\right).
\label{p7} 
\end{eqnarray}
For a central potential $V(r)$, this simplifies in the symmetric gauge 
(\ref{symgauge}) to
\begin{eqnarray}
H = \frac{\left({\bf{p}} - \bf{A}\right)^2}{2m} + V - \frac{\theta}{2r}\frac{\partial V}{\partial r}\left(L_z - \frac{B}{2}r^2\right).
\label{p8} 
\end{eqnarray}
Again for a harmonic potential $V(r) = \lambda r^2$, this takes the form 
\begin{eqnarray}
H = \frac{{\bf{p}^2}}{2m} + \frac{B^{\prime2}}{8m}r^2 - \tilde{\Lambda}L_z
\label{p9} 
\end{eqnarray}
where ${B^{\prime}} = B\sqrt{1 + \frac{8m\lambda}{B^2}(1 + \frac{\theta B}{2})}$ and $\tilde{\Lambda} = \frac{B}{2m} + \theta\lambda$.
Recognising that the structure of (\ref{p9}) is the same as that of (\ref{101d}),
we can readily write down the ground state wave-function as
%------------------------------
\begin{eqnarray}
\psi_{0}(z,\bar z; \theta) &=& \exp\left(-\frac{B^{\prime}(\theta)}{4}\bar{z} z
\right)\nonumber\\
 &=& \exp\left(-\frac{1}{4}\bar{z}z\sqrt{B^2 + 8m\lambda\left(1+\frac{\theta B}{2}\right)}\right)\;;\nonumber\\
 &&|\theta|<<1\;.
\label{p90} 
\end{eqnarray}

%------------------------------
This expression clearly reveals the fact that the commutative
wave-function has a non-trivial flow in $\theta$ of its own, only in the 
presence of interaction ($\lambda\neq0$) and the values of both
non-commutative wave-functions $\hat \psi$, $\hat \psi_{sw}$ and the commutative one $\psi$ 
coincide at $\theta=0$. One can, in principle, determine the exact
expression of this wave-function, valid upto all orders in $\theta$, 
but we shall not require this here. In fact the
wave-function (\ref{p90}) or higher angular
momentum states $z^l\psi_{0}(z,\bar z;\theta)$ can
be alternatively determined from perturbation theory
applied to each angular momentum sector $l$ for small
$\theta$ and $\lambda$. However, one point that we would
like to emphasise is that the Seiberg-Witten map does not map the non-commutative field 
$\hat\psi_{sw}(z,\bar z;\theta)$ at value $\theta$ to the corresponding one at the
commutative level $\psi(z, \bar z;\theta)$; the Seiberg-Witten map or equivalently the
Seiberg-Witten equation (\ref{swmap20}) only relates $\hat\psi(z, \bar z;\theta)$ to
$\hat\psi(z, \bar z;\theta=0)=\psi(z, \bar z;\theta=0)$.
 Furthermore, the fact that the
parameter $\hat m_{sw}(\theta)$ 
(\ref{z12})
 does not reproduce the expression to leading order in $\theta$, 
derived in \cite{chakraborty}  can be seen to follow from the observation that  the parameter $m$ was basically fixed by demanding the
 form invariance of the Schr\"{o}dinger action
which is equivalent to the stability analysis (in absence of interaction)
we have carried out in the previous sections. Also observe that in
 \cite{chakraborty} the ``renormalised" mass parameter $m$
 does not get modified by the interaction term in any way,
 in contrast to both $\hat m_{sw}$ (\ref{z12}) and $\hat m$ (\ref{600a}). 
On the other hand, the commutative wave-function $\psi$ in (\ref{p90}) 
gets modified in presence of interaction, as we mentioned 
above, in such a way that it has a non-trivial flow in $\theta$. 
This is in contrast to the non-commutative wave-functions $\hat\psi$ (\ref{z3})
 and $\hat\psi_{sw}$ (\ref{z5})
which have flows in $\theta$ even in absence of interactions.
 Finally, note that we have three versions of the Hamiltonians
 here with distinct transformations properties : (i) $\hat{H}$ occuring
 in (\ref{g1}) transforms adjointly under $U(1)_\star$ gauge transformation,
 (ii) $H$ 
occuring in  (\ref{p7}) transforms adjointly under ordinary $U(1)$ 
gauge transformation and (iii) the Bopp-shifted Hamiltonian 
$\hat{H}_{BS}$ occuring in (\ref{g1}) which, however, does not have any of these transformation properties under either type of gauge transformation as it was
constructed just by disentangling the $\star$ product but retaining 
the non-commutative variables. In this context, it will be worthwhile
 to remind the reader that in order to have the symmetry under 
$\star$ gauge transformation we must have non-commutative variables
composed through $\star$ product and to have the corresponding symmetry under ordinary gauge transformation, we must replace the non-commutative variables by commutative ones by making use of the Seiberg-Witten map apart from disentangling the $\star$ product as was done in \cite{chakraborty}. Consequently, the issue of maintaining the gauge invariance/covariance is not relevant here, since we are dealing with $\hat{H}_{BS}$ in this paper.

%%%%%%%%%%%%%%%%%%%%%%%%%%%%%%%%%%%%%%%%%%%%%%%%%%%%%%%%%%%%
%%%%%%%%%%                  Constructing dualities
%%%%%%%%%%%%%%%%%%%%%%%%%%%%%%%%%%%%%%%%%%%%%%%%%%%%%%%%%%%%
\section{Constructing dualities}
\label{duality}

We have seen how physically equivalent families
 of non-commuting Hamiltonians can be constructed.
 In this construction $\theta$ simply plays the role
 of a parameter and subsequently, as the physics does not change,
 physical quantities can be computed with any value of this parameter. A natural question to pose, therefore,
 is whether there is any advantage in choosing
 a specific value of $\theta$, i.e., is there any advantage
 in introducing non-commutativity in the first place. The motivation
 for asking this question was already outlined
 in section \ref{general}, where it was pointed
 out that in some existing literature \cite{jelal}
 the non-commutative quantum Hall system is considered
 a paradigm for the fractional quantum Hall effect which,
 however, requires the presence of interactions.
 If this interpretation is to be taken seriously
 a natural possibility that presents itself is that
 interacting commuting systems may in some approximation
 be equivalent to a particular non-interacting non-commutative system. If this turns out to be true, it would provide a
 new rational for the introduction of non-commutativity
 in quantum Hall systems.
 In this section we explore this possibility within
 a very simple setting.

We consider the non-commutative harmonic oscillator moving
 in a constant magnetic field discussed in section \ref{harmonic}.
 After undoing the star product through a Bopp-shift
 we find the Hamiltonian
\begin{eqnarray}
\hat H_{BS}(\theta) &=& \frac{{\bf p}^{2}}{2M_0} + \frac{{\bf x}^2}{2}M_0{\Omega_0}^{2} - \Omega_0(\theta)L_{z}\nonumber\\&&+\hat\lambda \left(\frac{\theta^2}{4}{\bf p}^{2}+{\bf x}^{2}-\theta L_z\right)\nonumber\\
&=& \hat H_0+\hat V
\label{dual1} 
\end{eqnarray}
where
\begin{eqnarray}
&&\frac{1}{2M_0} = \frac{\left(1 + \frac{\bar B\theta}{4}\right)^2}{2\hat m},\nonumber \\
&&\frac{1}{2}M_0{\Omega_0}^{2} = \frac{\bar B(\theta)^2}{8\hat m(\theta)}\;.
\label{dual2}
\end{eqnarray}
To represent equivalent systems, the parameters $\bar B$, $\hat m$ and $\hat\lambda$ are parameterized as in (\ref{s5}), (\ref{600a}) and (\ref{604a}), respectively.  

Naively one might argue that when the non-commutative coupling constant $\hat\lambda$ becomes small, the interaction term can be neglected on the non-commutative level.  However, as this happens when $\theta$ becomes large ($\hat\lambda\sim 1/\theta$), one sees from the Bopp-shifted equivalent of the Hamiltonian that this is not true due to the $\theta$ dependency that is generated by the Bopp-shift.  One therefore has to use a different criterion to decide when the interaction term $\hat V$ is small and can be neglected. One way is to introduce a norm on the space of operators and check that $\hat V$ is small in this norm. The trace norm ${\rm tr}(\hat V^\dagger \hat V)$ is divergent and cannot be used; a regularization is required.  An obvious alternative candidate to use is the following
\begin{eqnarray}
Z(\theta)=\frac{{\rm tr}(\hat V^\dagger e^{-\beta \hat H_0} \hat V)}{{\rm tr} e^{-\beta \hat H_0}}\;.
\label{dual3}
\end{eqnarray}

Here $\beta$ plays the role of an energy cutt-off.  It is clear that $Z(\theta)$ has all the properties of a norm, in particular $Z(\theta)=0$ if and only if $\hat V=0$. As remarked before, it is impossible to eliminate $\hat V$ completely, however, we can minimize $Z(\theta)$ with respect to $\theta$ and in doing this find the value of $\theta$ for which the  non-commutative non-interacting Hamiltonian $\hat H_0$ gives the best approximation to the interacting Hamiltonian. Since the low-energy spectrum of $\hat H_0$ is biased in the norm (\ref{dual3}), one can expect that the low-energy spectrum of $\hat H_0$ would give good agreement with the interacting spectrum, while the agreement will become worse as one moves up in the spectrum of $\hat H_0$. Before implementing this program, there is one further complication to take care of. Due to the degeneracy of $\hat H_0$ in the angular momentum, the norm (\ref{dual3}) is still divergent when summing over angular momenta in the trace. However, since $\hat V$ is a central potential and subsequently different angular momentum sectors decouple, it is quite sufficient to implement the program above in each angular momentum sector seperately.  Under minimization this will give rise to an angular momentum dependent value of $\theta$, giving rise to a lifting in the degeneracy in angular momentum, which is what one would expect in the presence of interactions.  To proceed we therefore replace (\ref{dual3}) by           
\begin{eqnarray}
Z(\theta,\ell)&=&\frac{{\rm tr}_\ell(\hat V^\dagger e^{-\beta \hat H_0} \hat V)}{{\rm tr}_\ell e^{-\beta \hat H_0}}\nonumber\\
&=&\sum_{n_-=0}^{\infty}|\langle n_-,\ell|V|n_-,\ell\rangle|^2e^{-\beta \Omega_0 (2n_-+1)}\;
\label{dual3a}
\end{eqnarray}
where ${\rm tr}_\ell$ denotes that the trace is taken over a fixed angular momentum sector, (\ref{10.1}) was used and $|n_-,\ell\rangle$ denote the eigenstates of $\hat H_0$.  This expression can be evaluated straightforwardly to yield
\begin{eqnarray}
Z(\theta,\ell)&=&\hat\lambda^2(\theta)\left[\Gamma(\theta)^2\left(1+\frac{2}{\sinh^2(\beta\Omega_0)}\right)\right.\nonumber\\
&+&2\ell\coth(\beta\Omega_0)\Gamma(\theta)\left(\Gamma(\theta)-\theta\right)\nonumber\\
&+&\left.\ell^2\left(\Gamma(\theta)-\theta\right)^2\right]\;,
\label{dual4}
\end{eqnarray}
where
\begin{eqnarray}
\Gamma(\theta)=\frac{M_0\Omega_0\theta^2}{4}+\frac{1}{M_0\Omega_0}\;.
\label{dual4a}
\end{eqnarray}
For $\beta>>1/B$ one finds the value of $\theta$ that minimizes this expression to be 
\begin{eqnarray}
\theta(\ell)=\frac{2(1+\ell)}{B(1+2\ell)}
\label{dual5}
\end{eqnarray}
at which value $Z(\theta,\ell)\sim\frac{1}{B^2}$, which means that the potential at these values of $\theta$ can be treated as a correction of order $1/B$.  The eigenvalues of $\hat H_0$ at these values of $\theta$ are easily evaluated to be 
\begin{eqnarray}
E_{n_-}(\ell)&=&2\Omega_0(\ell)(n_-+1/2)\;,\nonumber\\
\Omega_0(\ell)&=&\frac{B}{4m}\left(1+\sqrt{1+\frac{16\lambda m(\ell+1)}{B^2}}\right)\;.
\label{dual6}
\end{eqnarray}
From the above considerations it is clear that the approximation is controlled by $1/B$.  One therefore expects (\ref{dual6}) to agree with the exact result (\ref{z2}), at least for the lowest eigenvalues, to order $1/B$.  This indeed turns out to be the case.  Expanding the lowest eigenvalues of (\ref{dual6}) and (\ref{z2}) to leading order in $1/B$ one finds in both cases
\begin{eqnarray}
E_0(\ell)=\frac{B}{2m}+\frac{2(\ell+1)\lambda}{B}\;.
\label{dual7}
\end{eqnarray}
This result suggests that it is indeed possible to trade the interactions for non-commutativity, at least in the lowest Landau level and for weak Landau level mixing (large $B$). It would, of course, be exceedingly naive to immediately extrapolate from the above to realistic quantum Hall systems.  However, the above result does suggest a new paradigm for non-commutative quantum Hall systems worthwhile exploring. Within this paradigm interactions get traded, at least in the lowest Landau level, for non-commutativity, explaining the fractional filling fractions and emergence of composite fermions from a new perspective.  
%%%%%%%%%%%%%%%%%%%%%%%%%%%%%%%%%%%%%%%%%%%%%%%%%%%%%%%%%%%%%%%%%
%%%%%%%%%%%            Discussion and conclusions
%%%%%%%%%%%%%%%%%%%%%%%%%%%%%%%%%%%%%%%%%%%%%%%%%%%%%%%%%%
\section{Discussion and conclusions}
\label{conclusions}

We have demonstrated how physically equivalent families of non-commutative Hamiltonians can be constructed.  This program was explicitly implemented to all orders in the non-commutative parameter in the case of a free particle and harmonic oscillator moving in a constant magnetic field in two dimensions.  It was found that this spectrum preserving map coincides with the Seiberg-Witten map in the case of no interactions, but not in the presence of interactions. A new possible paradigm for non-commutative quantum Hall systems was demonstrated in a simple setting.  In this paradigm an interacting commutative system is traded for a weakly interacting non-commutative system, resulting in the same physics for the low energy sector.  This provides a new rational for the introduction of non-commutativity in quantum Hall systems.  

\section*{Acknowledgements}

This work was supported by a grant under the Indo--South African research
 agreement between the Department of Science and Technology,
Government of India and the South African National Research Foundation.
  FGS would like to thank the S.N. Bose National Center for
 Basic Sciences for their hospitality in the period that this 
work was completed. BC would like to thank the Institute of Theoretical
Physics, Stellenbosch University for their hospitality during the period when part of this work was completed.

\appendix
\section{Eigenvalues and eigenstates of the free and harmonic oscillator Hamiltonians}
\label{appendix}

To solve for the eigenvalues and eigenstates of (\ref{5}) one introduces creation and annihilation operators through the equations
\begin{eqnarray}
b_x = \sqrt{\frac{M\Omega}{2}}\left(x + \frac{ip_x}{M\Omega}\right),\quad   b_{x}^{\dagger} = \sqrt{\frac{M\Omega}{2}}\left(x - \frac{ip_x}{M\Omega}\right)\;,\nonumber\\
b_y = \sqrt{\frac{M\Omega}{2}}\left(y + \frac{ip_y}{M\Omega}\right),\quad  b_{y}^{\dagger} = \sqrt{\frac{M\Omega}{2}}\left(y - \frac{ip_y}{M\Omega}\right)\;.\nonumber\\
\label{6} 
\end{eqnarray}
In terms of these operators the Hamiltonian (\ref{5}) takes the form,
\begin{eqnarray}
 H = \Omega \left(b_{x}^{\dagger}b_x + b_{y}^{\dagger}b_{y} + 1\right) - i\Omega\left(b_{x}b_{y}^{\dagger} - b_{x}^{\dagger}b_{y}\right)\;.
\label{7} 
\end{eqnarray}

From these creation and annihilation operators, written in a ``cartesian basis", one can
define corresponding creation and annihilation operators in a ``circular basis" by making
use of the transformations
\begin{eqnarray}
b_{+} = \frac{1}{\sqrt{2}}\left(b_{x} - ib_{y}\right)\; , \quad   b_{+}^{\dagger} = \frac{1}{\sqrt{2}}\left(b_{x}^{\dagger} + ib_{y}^{\dagger} \right) \;,\nonumber\\
b_{-} = \frac{1}{\sqrt{2}}\left(b_{x} + ib_{y}\right)\;, \quad  b_{-}^{\dagger} = \frac{1}{\sqrt{2}}\left(b_{x}^{\dagger} - ib_{y}^{\dagger}\right)\;.
\label{9} 
\end{eqnarray}
Using these, the Hamiltonian (\ref{7}) becomes
\begin{eqnarray}
 H &=& \Omega \left(b_{+}^{\dagger}b_{+} + b_{-}^{\dagger}b_{-} + 1\right) - \Omega\left(b_{+}^{\dagger}b_{+} - b_{-}^{\dagger}b_{-}\right) \nonumber \\
 &=& \Omega \left(n_{+} + n_{-} + 1\right) - \Omega\left(n_{+} - n_{-}\right)\nonumber \\
&=& 2\Omega\left( n_{-} + \frac{1}{2}\right)\;.
\label{10} 
\end{eqnarray}
Note that the energy spectrum depends only on $n_{-}$ and is independent of $n_{+}$. Therefore, it results in an infinite degeneracy in the energy spectrum.
 The above cancellation of the terms involving $n_{+}$ has taken place since
 the coefficients of $n_{+}$ are equal. This is also true
 in the limit $\theta = 0$. This feature does not persist in presence
of interactions (see section 4).  

Introducing complex coordinates $z = x + iy$ and $\bar z = x - iy$, 
(\ref{9}) takes the form
\begin{eqnarray}
b_{+} = \frac{1}{2}\sqrt{M\Omega}\left[\bar z + \frac{2}{M\Omega}\partial_{z}\right]\;, \   b_{+}^{\dagger} = \frac{1}{2}\sqrt{M\Omega}\left[z - \frac{2}{M\Omega}\partial_{\bar z}\right]\;,\nonumber\\
b_{-} = \frac{1}{2}\sqrt{M\Omega}\left[z + \frac{2}{M\Omega}\partial_{\bar z}\right]\;, \   b_{-}^{\dagger} = \frac{1}{2}\sqrt{M\Omega}\left[\bar z - \frac{2}{M\Omega}\partial_{z}\right]\;.\nonumber\\
\label{10b} 
\end{eqnarray}
The ground state wave-function is annihilated by $b_{-}$, i.e. $b_{-}\hat\psi(z, \bar z; \theta) = 0$. This immediately leads to the solution
\begin{eqnarray}
\hat\psi_{0}(z, \bar z;\theta) = N\exp\left[-\frac{M\Omega}{2}\bar{z}z\right]
 =  N\exp\left[- \frac{\bar B}{4\left(1 + \frac{\bar B\theta}{4}\right)}\bar{z}z\right].\nonumber\\
\label{10c} 
\end{eqnarray}
Since $\bar B(\theta = 0) = B$,  the above solution goes smoothly
 to the commutative result
\begin{eqnarray}
\psi(z, \bar z) = N\exp\left[- \frac{B}{4}\bar{z}z\right]\;.
\label{10d} 
\end{eqnarray}
This state is also annihilated by $b_{+}$ and
 therefore corresponds to zero angular momentum state, as the angular momentum 
operator $L_{3}=(xp_{y}-yp_{x})$ takes the following form in cartesian
or circular basis
\begin{eqnarray}
L_{3}=i\left(b_{x}b_{y}^{\dag}-b_{x}^{\dag}b_{y}\right)
=\left(b_{+}^{\dag}b_{+}-b_{-}^{\dag}b_{-}\right).
\label{1000a} 
\end{eqnarray}
If this $xy$-plane is thought to be embedded in $3-d$ Euclidean space
$\mathcal{R}^3$, then the other rotational generators $L_{1}$ and  $L_{2}$
obtained by cyclic permutation would result in the standard angular
momentum $SU(2)$ algebra 
\begin{eqnarray}
\left[L_{i}, L_{j}\right]=i\epsilon_{ijk}L_{k}.
\label{1000b} 
\end{eqnarray}
One can, however, define the $SU(2)$ algebra using the creation and annihilation operators alone, which, in the cartesian basis, is given by
\begin{eqnarray}
J_{1} &=& \frac{1}{2}\left(b^{\dagger}_{x}b_{x} - b^{\dagger}_{y}b_{y}\right)\;, \nonumber \\
J_{2} &=& \frac{1}{2}\left(b^{\dagger}_{x}b_{y} + b^{\dagger}_{y}b_{x}\right)\;, \nonumber \\
J_{3} &=& \frac{1}{2i}\left(b^{\dagger}_{x}b_{y} - b^{\dagger}_{y}b_{x}\right)\;,
\label{8} 
\end{eqnarray}
satisfying $\left[J_{i}, J_{j}\right]=i\epsilon_{ijk}J_{k}$. As one can
easily verify, by computing the Poisson brackets of the generators
with phase-space variables that $J_{1}$ generates
rotation in $(x, p_{x})$ and $(y, p_{y})$ planes, $J_{y}$ in $(x, p_{y})$
and $(y, p_{x})$ planes and $J_{z}$ in $(x, y)$ and $(p_{x}, p_{y})$ planes.
Also note that $L_{3}$ is not identical to $J_{3}$ but differs by a factor
of $2$: $L_{3}=2J_{3}$.

The Casimir operator in terms of $J_{i}$ representation now becomes
\begin{eqnarray}
\vec{J}^2= \frac{1}{4}\left(b_{+}^\dag b_{+}+b_{-}^\dag b_{-}\right)
\left(b_{+}^\dag b_{+}+b_{-}^\dag b_{-}+2\right).
\label{840} 
\end{eqnarray}
with eigenvalues
 $\vec{J}^2= \frac{1}{4}\left(n_{+}+n_{-}\right)\left(n_{+}+n_{-}+2\right).$
Defining $n_{+}+n_{-}=2j$, the Casimir becomes $\vec{J}^2=j(j+1)$. Also,
if the eigenvalues of $J_{3}$ is given by $l^{\prime}$, then the
eigenvalues of $L_{3}$ will be given by $n_{+}-n_{-}=2l^{\prime}=l\epsilon 
\mathcal{Z}$. Note that, like $l^{\prime}$, $j$ also admits half-integral
values. Finally, one can write down the eigenvalues (\ref{10}) as
\begin{eqnarray}
E_{n_{-}}=\Omega\left(2j-2l^{\prime}+1\right)=\Omega\left(2j-l+1\right)
\label{841} 
\end{eqnarray}
which agrees with \cite{gamboa}. Any arbitrary state can now be
obtained by repeated application of $b_{\pm}^{\dag}$ on (\ref{10c}) as
\begin{eqnarray}
\vert n_{-}, l\rangle \sim \left(b_{-}^{\dag}\right)^{n_{-}}
\left(b_{+}^{\dag}\right)^{l}\hat\psi_{0}(z, \bar z;\theta)\;.
%\sim
%\left(\bar{z} z\right)^{n_{-}}z^l\hat\psi_{0}(z, \bar z;\theta).
\label{842} 
\end{eqnarray}
 
 %%%%%%%%%%%%%%%%%%%%%%%%%%%%%%%%%%%%%%%%%%%%%%%%%%%%%%%%%%%%%%%%%%%%%%%%%%%%%%%%%%%%%%%%%%%%%%%%%%%%%%%%%%%%%%%

\end{document}